%Paper: astro-ph/9506105
%From: Basilio Santiago <santiago@ast.cam.ac.uk>
%Date: Tue, 20 Jun 1995 16:11:40 +0100 (BST)

\magnification=\magstep1
\hoffset=-0.6 true cm
\voffset=0.0 true cm
\baselineskip=20pt minus 1pt
\baselineskip=12pt

\vsize=8.9truein
\hsize=6.8truein
%\hyphenpenalty=10000
%\raggedright
%\parskip=\medskipamount
\tolerance=10000
\parindent=1truecm
\raggedbottom
\def\pp{\parshape 2 0truecm 5.8truein 2truecm 5.01truein}
\def\ltsima{$\; \buildrel < \over \sim \;$}
\def\simlt{\lower.5ex\hbox{\ltsima}}
\def\gtsima{$\; \buildrel > \over \sim \;$}
\def\simgt{\lower.5ex\hbox{\gtsima}}
\def\bline{\hbox to 1 in{\hrulefill}}
\def\etal{{\sl et al.\ }}

\def\vs{{\it vs~}}

\def\kms{\ifmmode {\rm \ km \ s^{-1}}
\else
$\rm km \ s^{-1}$\fi}

\def\vp{\ifmmode {\bf r}
\else
${\bf r}$\fi}
\def\lumf{\ifmmode {\Phi ({\rm L})}
\else
{$\Phi ({L})$}\fi}
\def\lfvp{\ifmmode {\Phi ({\rm L,{\bf r}})}
\else
{$\Phi ({L,{\bf r}})~$}\fi}
\def\ab{\ifmmode {A_B}
\else
${A_B}$\fi}
\def\ablb{\ifmmode {A_B (l,b)}
\else
${A_B (l,b)}$\fi}
\def\mlim{\ifmmode {\rm m_{lim}~}
\else
${m_{lim}~}$\fi}
%\nopagenumbers
%\headline{\hfill \folio \hfill}

\centerline{\bf HST STAR COUNTS AT HIGH GALACTIC LATITUDES}
\bigskip
\centerline{Bas\'\i lio X. Santiago, Gerard Gilmore, Rebecca A. W. Elson}
\vskip 0.5 true cm
\centerline {Institute of Astronomy,
Cambridge University, Madingley Road, Cambridge CB3 0HA, United
Kingdom}
\bigskip
\centerline{\it Submitted to MNRAS}
\bigskip
\centerline{\bf Key words: Galaxy: stellar content and structure, Stars:
low-mass and statistics}
\bigskip
\bigskip
{\bf ABSTRACT}
\medskip
We use star counts from 13 deep HST fields imaged with the {\it Wide
Field Camera - 2} in order to constrain the amount of dark matter in the
Galaxy that can be in the form of low-mass main sequence stars or
white-dwarfs. Based on the number of red stars found in our fields, we
exclude the possibility
that more than 25\% of the massive dark halo is made up of M dwarfs
or subdwarfs;
fairly massive ($M \sim 0.2 M_\odot$)
and yet extremely faint ($M_I$ \gtsima $14.5$)
stellar candidates would have to be invoked in order to make
the observed number of stars compatible with that predicted by
a stellar dark halo. White dwarfs must also be intrinsically
very faint ($M_I$ \gtsima $14$) in order to be consistent with the
observed number of faint stars in the HST fields.
We also rule out an increasing or flat
stellar luminosity function beyond $M_V \sim 13$. The inferred slope
of the disk luminosity function is intermediary
between local, volume-limited surveys and
ground-based photometric ones.
Finally, the magnitude counts are well fitted by existing
models for the structure of the Galaxy, with only small changes in the
fiducial values of the model parameters. The colour distribution, however, is
not well reproduced by the models. It is unclear at present if
this reflects inadequacies of the available models or
uncertainties in the colour-magnitude
diagrams for low metallicity stars and in the photometric calibration.
\vfill\eject
{\bf 1 INTRODUCTION}
\medskip
The dynamically inferred mass of the Galaxy seems to exceed
current estimates
of the contribution by stars, gas and dust
(Fich \& Tremaine 1991).
Evidence for non-luminous matter
exists in other galaxies, particularly large spirals with well
determined rotational velocity curves (Faber \& Gallagher 1979,
Rubin, Ford \& Thonnard 1982).
Contrary to the other, cosmologically more relevant,
``missing mass problems'',
the amount of Galactic dark matter
can still comfortably lie within the upper limit of baryonic
matter imposed by the Standard Big Bang nucleosynthesis
model (Silk 1990, Carr 1994,
Walker \etal 1991, Mathews, Schramm \& Meyer 1993).
Thus, it is certainly wise to consider low-mass stars or gas
as possible candidates for the Galactic missing matter.

In recent years, some evidence has been found for a steeply rising
stellar luminosity function in several of the
Halo globular clusters (Richer \etal 1991,
Richer \& Fahlman 1992). Despite the
uncertainties in the mass-luminosity relation for low-mass, low-metallicity
stars (cf., Elson \etal 1995), these results have suggested
that the mass of the
stellar populations in the halo of the Galaxy
may be larger than previously thought.
However, ground-based studies of globular clusters are severely affected by
crowding and thus necessarily involve large completeness corrections.
Similarly, field star-count studies and searches for
new stellar populations, specially in the Galactic spheroidal component,
are seriously limited by the difficulty of
separating stars from extragalactic objects at magnitudes beyond $V \sim 20$
(Reid \& Gilmore 1982, Stobie \& Ishida 1987, Reid 1990, Kron 1980).
This limitation prevents strong observational constraints being
placed on the number of low-mass, main-sequence, stars with $M_V$ \gtsima
$13$, since such objects are probed only within fairly local volumes,
out to distances of no more than 250 pc. Thus, besides being
hampered by small number
statistics, ground-based field star counts of low luminosity stars do not
reach distances beyond one disk scale height, even in high-latitude fields.
Another important issue for such studies is the existence of
selection effects caused by photometrically unresolved binaries, leading to
a possible confusion in the conversion from observed to true star counts.

Much progress on both globular cluster and field star count studies
can be achieved with the {\it Hubble Space Telescope (HST)}.
Recent globular cluster faint star counts have been carried out by
Paresce, De Marchi \& Romaniello (1995) and
Elson \etal (1995), requiring relatively
small completeness corrections. No evidence for a steeply rising
luminosity function was found.
In the field, with HST's much improved spatial resolution,
star-galaxy separation
can be reliably performed down to $V \sim 24-26$, allowing
us to probe the faint end of both disk and halo main sequences
within distances of about 1-2 kpc.
The larger depth achievable by the HST field
star counts opens up the possibility of further
constraning the current models for the structure of the Galaxy (Bahcall
\& Soneira 1980,1984; Gilmore \& Reid 1983, Reid \& Majewski 1993).
Due to the
difficulty in probing well into the spheroidal component,
relatively poor constraints have been obtained so far for the
luminosity function and density profiles of the Galaxy's stellar halo.
Another poorly defined region in
parameter space describing the structure of the Milky Way is the
intermediate Population II component, or thick disk,
though considerable progress has recently been made in quantifying its
properties (Ojha \etal 1994).

In this paper we present star counts for 13 deep {\it Wide Field Camera - 2}
(WFPC-2) high-galactic latitude
fields. In \S II we describe our data, the
reduction and calibration methods and
the stellar sample selection. We also discuss the issues of
star-galaxy separation and sample completeness corrections.
In \S III we show our results.
We compare the observed number of red stars with predictions
based on the assumption that the dark halo of the Galaxy is made up of
such objects. The case for white-dwarfs as a dark matter
candidate is discussed as well.
We use the counts of red stars to constrain the slope of the faint end of
the stellar luminosity function and compare it with previous estimates.
Finally, we compare the entire observed magnitude and color distributions
with the predictions made by existing models of Galactic structure.
Our conclusions are presented in \S IV.
\bigskip
{\bf 2 THE STELLAR SAMPLE}
\medskip\nobreak
{\bf 2.1 The data and sample selection}
\medskip\nobreak
The 13 WFPC-2 fields used in this work were obtained as part of the
Medium Deep Survey key HST project (MDS). They are all fairly deep,
high-galactic latitude fields as can be assessed from Table~1.
Column 1 lists the MDS field identification name, columns 2 and 3 give their
galactic coordinates and columns 4 and 5 list the exposure times in both
HST I and V wide filters (F814W and F606W, respectively).
Each field covers an area of 5.7 $arcmin^{2}$.
All MDS fields are the result of parallel observations
made during the HST General Observing time. For more information
about MDS itself and its many applications we refer the reader to
Griffiths \etal (1994).

All fields consisted of multiple exposures which were
put through the standard HST pipeline reduction method (Holtzmann \etal 1995).
This consists of overscan, residual bias and dark current subtraction,
flat-fielding, and correction for several instrumental effects.
The individual calibrated frames were then
coadded and median-filtered in order to eliminate cosmic ray
events. For that purpose, use was made of an IRAF package written
by K. Glazebrook.
Some fields were
dithered, so that hot pixels were also eliminated during the stacking
process. The others had the hot pixels identified and masked
out using software kindly provided by N. Tanvir.
The number of
individual frames available in each field and for each bandpass
is listed in columns 6 and 7 of Table~1.

Object detection was carried out using the APM software
IMAGES. We adopted a
detection threshold of $1.5*\sigma_{bkg}$, where $\sigma_{bkg}$ is the
standard deviation in the background counts for each field and WFPC-2
chip. We detected objects in both F814W and F606W frames and defined as our
primary object list those objects common to both.
This helped eliminate cosmic ray residuals from our sample.
The adopted threshold of detection corresponded to
surface brightness levels in the ranges
$23.0$ \ltsima $\mu_{F814W}$ \ltsima $25.0~mag~arcsec^{-2}$ and $24.0$ \ltsima
$\mu_{F606W}$ \ltsima $26.0~mag~arcsec^{-2}$ for the HST I and V bands,
respectively.
The output files of IMAGES list positions,
shape parameters, magnitudes and a central intensity to total flux
ratio for each object detected, $I_{peak}/F_{tot}$.
In figure 1 we show the plot of this latter quantity as a function of
HST I band magnitude, $I_{F814}$, for the objects
in the 3 WFC-2 CCD chips in one
of our fields (uj70).

Two main loci of objects are clearly discernible from figure 1: one with
$I_{peak}/F_{tot} \sim 0.2-0.3$, spanning most of the
magnitude range and the second with increasing  $I_{peak}/F_{tot}$
towards fainter magnitudes.
The bright objects in the upper clump in figure 1
have $I_{peak}/F_{tot}$ values and
light profile characteristics (shape, FWHM, etc)
which resemble those
expected for point sources (WFPC-2 Instrument Handbook version 2.0).
We thus identify this locus as being predominantly made up
of stars. In fact, all the obvious galaxies in the chips
lie in the lower locus.

The two loci merge at both bright and faint magnitudes.
At the bright end, the declining trend of the stellar locus
is caused by saturation. At faint enough magnitudes, only the unresolved
center of
extended sources lies above the background noise and it is then
impossible to separate them from point sources.
We should also point out that most spurious detections such as
substructure in the PSF of bright stars also lie in the lower
locus. Thus, we use the diagram shown
in figure 1 to define a range in magnitudes
within which we can reliably select a sample of unsaturated stars.
The bright end cut-off value was typically within the interval
$18 < I_{F814} < 19$, whereas the faint end limit for reliable
classification was around $I_{F814} = 23.5-25.0$. Between
the bright and faint cut-off values,
the stellar sample was then defined by applying the star-galaxy classifier
defined as the solid line in figure 1.
All fields were visually inspected in search of contaminating non-stellar
objects wrongly included in the sample. No significant number of obvious
non-stellar objects was found.

The other 12 fields do not have as many stars as the one shown and their
upper locus in the $I_{peak}/F_{tot}$ \vs $I_{F814}$ diagram is
correspondingly less populated.
However, our experience with low galactic latitude data shows that
the position of the two loci is quite stable from field to field.
Thus, the same classifier was applied for all the 13 fields.

Throughout this work we
have restricted ourselves only to the 3 WFC chips, since the PC chip, besides
covering only a small area in the sky, would require a different star-galaxy
classifier. Also, we have not applied any correction for Galactic absorption
to the magnitudes and
colors; all the magnitude values quoted correspond to
uncorrected data. However, according to a standard Galactic
(cosec) reddening distribution, the maximum value
of $A_I$ among our fields is
$0.13~mag$ ($A_V \sim 0.23~mag$) for sources outside the Galaxy.

Our choice of classifier is arbitrary and
requires an objective way of assessing issues
like completeness of the stellar sample and its contamination by
extragalactic sources. This is done in the next section.

\medskip\nobreak
{\bf 2.2 Completeness corrections}
\medskip\nobreak

We tested the adopted star-galaxy classifier by means of simulations
of stars and galaxies. We used the IRAF ARTDATA package for that purpose.
Images of stars and galaxies were built from bright
templates using the MKOBJECTS task. The templates were selected from the
MDS fields themselves. 42 stellar and galactic high signal-to-noise
objects were used as templates. The galaxy templates spanned a wide variety
of morphologies and structural features.
Nearly 5000 fainter and more distant versions of both stars and galaxies
were created on dozens of artificial HST images, with variable
background noise. For the galaxies, the effects of
K-corrections and surface brightness
dimming were incorporated. This requires a redshift to be assigned to them.
Since redshifts have not yet been measured for most of
the galaxy templates, a value
was randonly chosen for each of them from the probability distribution
$P(z/I_{F814})$
of having a redshift $z$ given the observed HST I magnitude $I_{F814}$.
A standard
Schechter galaxy luminosity function, uniform density,
no-evolution model with $q_0 = 0.5$ was
assumed in estimating this probability.

A poissonian noise distribution was assumed for both the background
and object fluxes. Each artificial frame had about 200 objects, which is
typical of MDS data. Object detection was carried over the simulated data
using the same software and parameters as for the real data. The
$I_{peak}/F_{tot}~vs.~I_{F814}$ diagram for one of the simulations is
shown in figure 2. Notice that we successfuly reproduce the two loci
seen in figure 1. The somewhat discretized distribution of simulated points
in figure 2 is an artificiality caused by the finite number of templates.
{}From our simulated images we could then extract a
completeness function for the stars that incorporates the loss
of objects both during detection and due to the adopted classifier.
This completeness function is shown for different signal-to-noise
levels in figure 3, as indicated in the captions.
As expected, lower signal-to-noise data will tend to be less complete at
any given magnitude.
The three simulations shown span the range of
signal-to-noise observed in the data. Each data field was matched to a
simulation with similar noise level and the stars in that field
were assigned a weight
given by the inverse of the corresponding
completeness function in all the analysis shown
this paper.
Notice that the corrections are
small, our stellar sample being 60\% complete down to I=24 even
in the worst case. Given the fairly small amplitude of the completeness
corrections, their associated uncertainties caused by
possible differences between the simulations and the real data are
likely to be of secondary importance.
\bigskip

\medskip\nobreak
{\bf 2.3 Aperture Photometry and Photometric Calibration}
\medskip\nobreak

We performed aperture photometry for the stars selected in each field using
the APPHOT.PHOT IRAF task. We adopted a small aperture size of 2 pixel radius
(0.2'') in order to avoid contamination by remaining cosmic ray residuals
and hot pixels to the derived magnitudes. An aperture correction of
0.3 mag was necessary to account for
the light lying outside the adopted aperture (WFPC-2 Instrument Handbook
version 2.0).
We derived both HST I and V magnitudes, from which
an HST V-I colour followed automatically.
The instrumental magnitudes were then converted to the
standard Johnson-Cousins system using the transformations as given by
Holtzmann \etal (1995) and Harris \etal (1991).
The first paper converts from the WFPC-2 flight
system to the WFPC ground-based
system, defined by Harris \etal (1991). We then use the
relations given by these later authors in order to derive
magnitudes and colours in the
standard Johnson-Cousins system.
For the V band, this latter conversion includes a (B-V) colour term.
(B-V) colours were obtained from (V-I) using the relation given by
Prosser (1992).
Since photometric calibration may be
uncertain, specially in the blue and red ends of the stellar colour
distribution, we checked our results with an alternative
calibration method, using the transformations
given by Bahcall \etal (1994).

The two calibration methods agree very well in the I band: Bahcall \etal
I magnitudes are systematically fainter by 0.04 mag at the most. In the V band,
however, the alternative method leads to brighter (fainter) magnitudes for
blue (red) stars. The discrepancy is as large as $0.20 mag$ in the
extreme blue end and $0.50$ mag in the extreme red. Such differences are
worrying and in part stem from the large difference between
the HST F606W
and the standard V filters. Furthermore, (V-I) is a strong function of
the effective temperature for low-mass main-sequence stars. We thus refrain
{}from using the observed
differential colour counts in both red and blue extremes
for any purpose in the next sections. We will rely only
on integral numbers, as given by the standard calibration method
proposed by the WFPC-2 team. The effect of adopting the alternative photometric
transformations of Bahcall \etal (1994) will also be assessed in each section.

In figure 4 we show the colour magnitude diagram (CMD)
for our stellar sample.
It contains a total of 333 objects. With completeness corrections,
this number increases by 12\%.
Notice that the red end of the CMD spans a large range of apparent
magnitudes, indicating the presence of both disk and halo red dwarfs.
One can also distinguish two main although very wide clumps,
one clustered within the region around $ V-I \sim 1.5$
and the other seen at redder colours ($2$ \ltsima $V-I$ \ltsima $3$).
The first corresponds to halo stars close to the main-sequence turn-off
whereas the second is dominated by thick disk and
disk G and K dwarfs (Reid 1992).

\bigskip
{\bf 3 STAR COUNTS}
\medskip\nobreak
{\bf 3.1 Limits on the stellar dark matter}
\medskip\nobreak
In this section the observed numbers of red low-mass main sequence
stars are compared to model predictions under the assumption that the massive
halo of our Galaxy is made up of such objects. We use the
dark matter halo model proposed by Bahcall, Schmidt \& Soneira
(1983) to quantify the mass distribution in the Halo.
This model assumes a spherical dark matter
distribution with mass density profile given by
$$\rho (r) \propto { \bigl( { {r} \over {r_0} }  \bigr) }^{-1.8}, \eqno (1)$$
where $r_0 = 2 kpc$ and whose normalization is given by a mass density
of $0.009~M_\odot~pc^{-3}$ at the solar distance.
The gravitational field generated by adding this model dark matter component
to the distribution of luminous material provides a good fit to the observed
rotational velocity profile of the Galaxy.
Once the dark matter distribution is specified, the
expected number of stars which would be seen in our 13 HST fields
depends only on the absolute magnitude $M_I$ and mass M of the dark matter
stellar candidate. The first quantity defines
the maximum distance (and hence the
volume) out to which the candidate stars can be detected in each field.
To be consistent with the observed star counts, the effect of
extinction was incorporated
when estimating this maximum distance.
The predicted number of stars is then simply the
total dark halo mass within this volume divided by M.

In figure 5a we show these numbers for several values
values $M_I$ and M. The upper curve utilizes masses taken
{}from Kroupa, Tout \& Gilmore (1993). These authors derive a $M-M_V$ relation.
V band absolute magnitudes were taken from the $M_I$ values of
each candidate under the assumption that
$V-I = 3.0$ in all cases. This actually tends to underestimate
the number of halo stars, since the true colours should be redder than
$V-I=3.0$ for $M_I$ \gtsima $11.5$, leading to fainter $M_V$
and, consequently, smaller masses.
The lower curve makes the ultra-conservative assumption of
$M=0.2 M_\odot$ for all candidates, regardless
of their luminosity. Finally, the solid horizontal line gives the observed
numbers of HST stars with $V-I > 3.0$.
The dotted line represents
the 99 percentile position of a Poisson distribution with mean value
given by the observed number. We conclude that most faint stellar
candidates are ruled out at more than 99\% confidence level. In order to
reproduce the observed numbers one needs to invoke stars with $M_I$ \gtsima
$14.5$ but yet fairly massive. This violates what we know
about the mass-luminosity relation for main sequence stars, since
$M_I=14.5$ should be very close to the hydrogen-burning limit
($\sim 0.08 M_\odot$).

Another stellar dark matter candidate could be white-dwarfs (Larson
1986, Tamanaha \etal 1990).
In figure 5b we show the predicted number of white-dwarfs
as a function of $M_I$. We assume that $M = 0.65 M_\odot$, independently
of the absolute magnitude.
This value is well bracketed by the mass range found
in several different studies (Liebert 1980). The observed
number (and its corresponding
99\% Poisson deviate) now corresponds to objects in the blue end of the
colour magnitude diagram shown in figure 4 ($V-I < 0.8$). Again, very faint
($M_I$ \gtsima $14$) white-dwarfs are required if they are to account for
the entirety of the dark matter in the Milky Way. It is an interesting
coincidence that this luminosity limit is comparable to that of
the lowest-luminosity (disk) white dwarfs yet known (Liebert, Dahn \& Monet
1988). Current cooling models provide an age for such cool degenerates of about
10 Gyr (Wood 1992, cf von Hippel, Gilmore \& Jones 1995). Of course,
any white dwarf descendants of an early halo population will
have an age of that population. Such stars will have $M_{bol} (\sim M_I)
\sim 16$ (cf figure 3 of von Hippel \etal 1995), about
one magnitude fainter than our present limits. We note, however, that any
stars with mass initially below $\sim 1 M_\odot$ which might have formed
would have been visible in this survey.

There are two reasons to believe that the derived
constraints on the stellar dark mass
candidates are actually stronger than the ones inferred from figure 5.
First, not all objects
with $V-I > 3.0$ (or $V-I < 0.8$) are actually stars. Both red and blue stellar
counts may be contaminated by some unresolved very faint galaxies.
The blue sample should also be contaminated by quasars.
According to the simulations described in \S2.2, however, galaxy
contamination should be
smaller than 15\% at the faintest magnitude limits used in selecting our
sample, even for the shallowest fields.
Second, in comparing the predicted and observed numbers we are
implicitly assuming that the latter is made of halo stars only;
obviously, both red dwarf and white dwarf samples should be contaminated
by disk and thick disk stars. In fact,
models for the Galaxy predict that most of the faint red stars
we observe actually belong to these latter components (see \S 3.3).

We should also point out that, although there would be a change in the
number of stars redder than $(V-I) = 3.0$, the same conclusions would
apply had we used the alternative photometric calibration of Bahcall \etal
(1994).

An alternative way of
estimating the contribution of red subdwarfs
to the total Galactic mass is by constraining
the slope of the faint end of the halo luminosity function.
More specifically, we can compare the inferred value for this
slope, derived under the assumption that faint stars
account for the Galaxy's missing mass,
to the slope obtained from star counts constraints.

For a given domain in absolute visual magnitudes $M_{br} < M_V < M_{ft}$, we
assume that the halo stellar luminosity function, $\Phi_h (M_V)$,
is a power-law,
$$ \Phi_h (M_V) \propto 10^{0.4~\gamma_h~M_V},~M_{br} < M_V < M_{ft}~,
\eqno (2)$$
where $\gamma_h$ is the slope. We
normalize $\Phi_h (M_V)$ locally at the bright end of the specified domain
($M_{br}$) to 1/500 of the observed disk luminosity function, $\Phi_d (M_V)$
(Bahcall \& Soneira 1984).
Since this latter is known with reasonable
accuracy only down to $M_V \sim 13$, we adopt $M_{br} = 13$
as the normalization point. We also assume expression (2) to
be valid down to $M_{ft} = 19$, which should be very close to the hydrogen
burning limit.
Adopting the local normalization
for the mass density of the dark halo as given by Bahcall \etal (1983)
($\rho = 0.009 M_\odot pc^{-3}$)
and demanding it to be in the form of low-mass stars with $13 < M_V < 19$,
and with a mass-luminosity relation
given by Kroupa \etal (1993), we derive a slope of
$\gamma_h = 1.56$. Assuming $M=0.2 M_\odot$ for all faint stars leads to
$\gamma_h = 1.35$. These results are summarized in the first two lines
of Table~2. We list the nature of the constraint on $\gamma_h$,
the assumed range
in $M_V$, the adopted mass-luminosity relation
and the inferred slope of $\Phi_h (M_V)$.

These values for $\gamma$ can then be compared with that needed to
match the observed number of stars with $V-I > 2.40$ ($M_V > 13$, $M_I > 10.6$)
in our MDS fields. We again assume very conservatively
that all such stars belong to the halo of the Galaxy, with the same density
profile and luminosity function used in the previous experiment, even though
most of these stars are expected to be thick disk or disk stars.
We also use the same normalization at the bright end of the luminosity
function.
Notice that, for a given slope $\gamma_h$, the predicted number of stars
depends on the assumed range of $M_I$,
$10.6 < M_I < 14.1$, since the sample selection was based on the F814W
magnitudes. This range of $M_I$ was obtained from the $M_V(V-I)$ relation
given by Bahcall etal (1994) for spheroid and halo stars.
We again take extinction into account in our
computations.
We obtain $\gamma = 1.19$.
This value falls very short of those needed to fill in the Galaxy with
stellar dark matter, even if $M = 0.2 M_\odot$.
The results are listed in the second half of Table~2. In the last column,
we list the fraction of the dark matter that would correspond to the given
slope, for both Kroupa \etal (1993) (first value) and $M=0.2 M_\odot$ masses
(second value).
Notice that if we tried to account for the contribution
of disk stars, the slope obtained from the star counts (and consequently
the associated mass fractions) would
be yet smaller.
We should also point out the slope value
depends slightly on the interval in $M_V$ within which they are
fitted but the constraints on the amount of stellar dark matter
do not.
The same argument applies to the normalization used for the
halo luminosity function at $M_{br}$;
a value of 1/800 has been used in several star-count
studies (Bahcall \& Soneira 1980, Gilmore 1984), and is favoured
by most recent star-count analysis of Galactic structure (eg Soubiran
1993, Perrin \etal 1995). Had we adopted this value,
the slopes
would be larger in {\it both} experiments (dark matter and star counts
constraints), bearing no effect on our main conclusion.
Finally, the results based on the inferred value of $\gamma_h$
are again not significantly affected by photometric
calibration problems.

Thus, even under the very conservative
assumptions that stars as faint as $M_V=19$ have $M=0.2 M_\odot$ and that
all MDS red stars are in the halo,
the fraction of the dark halo mass due to the M subdwarfs
cannot exceed 50-60\%. Under more realistic assumptions on the
stellar masses,
we derive an upper limit of about 25\%. We note that this is in very good
agreement with recent deep recent infra-red surveys,
which derive tight upper limits on the number of faint halo
M dwarfs (Hu \etal 1994).

\medskip\nobreak
{\bf 3.2 The stellar luminosity function}
\medskip\nobreak

We now consider what constraints the observed counts of red stars impose
on the faint end of the stellar luminosity function under the more realistic
assumption that both disk and halo stars are contributing to these
counts. For the disk we again consider the slope ($\gamma_d$) in the interval
$13 \leq M_V \leq 19$, which corresponds to
$10.1 \leq M_I < 14.2$. Thus, $\gamma_d$ is determined by matching
model predictions to the observed counts with
$V-I \geq 2.9$. The corresponding range for $\gamma_h$ is
then $14.3 \leq M_V \leq 19.0$ ($11.4 \leq M_I \leq 14.1$).
These magnitude and colour intervals
assume $M_V(V-I)$ relations taken from Monet \etal (1992)
and Bahcall \etal (1994).
We use standard density laws for the thin disk and stellar halo;
the former has an exponential profile with a scale length of 3.5 kpc
and a scale height of 325 pc
for faint main sequence stars. For the stellar halo we adopt a de Vaucouleurs
profile with effective radius $r_e = 2700 pc$.
In Table~3, we list the results; column 1 gives the model components
whose contributions to star counts were matched to the observed numbers.
Column 2 lists the assumed
number of stars redder than $V-I = 2.9$ and
column 3 lists
the derived slope at the faint end.
If we assume the same power-law shape and slope for both disk and halo
luminosity functions,
we derive $\gamma_d = \gamma_h = -0.29$.
Assuming that all stars belong to the
disk leads to $\gamma_d = -0.26$.
We have also derived the slope values that match the 99\% Poisson
deviates of the observed numbers (last two lines).
Even if we overlook the contribution by halo stars,
we still obtain $\gamma_d = -0.09$.
Thus, our data are
inconsistent with a flat faint end for the stellar luminosity function
with more than 99\% confidence.
Also notice the
relatively small effect of including
the halo in the expected counts.
The main factor influencing the derived slope values
is probably the uncertainty in the MDS red star counts due to photometric
calibration. This uncertainty, however, is not large enough to
affect the conclusion of a decreasing luminosity function.
Other factors influencing the slope include
uncertainties in the CMD of low-mass stars (which
affect the observed numbers through the $V-I$ colour range associated
with $M_V > 13$) and
the $M_V$ interval used for fitting $\gamma$.

Our best estimate of the disk luminosity function
based on the star counts is shown as the solid line
in figure 6, along with the results of previous works (Wielen \etal 1983,
Jahreiss 1987, Dahn \etal 1986, Stobie, Ishida \& Peacock 1989;
see Bessel \& Stringfellow 1993).
It represents the derived value of the disk+halo luminosity function
slope ($\gamma_d=\gamma_h=-0.29$).
The two dashed lines correspond to the extreme cases of
using the 1\% and
99\% Poisson deviates of the observed number counts
and the most conservative assumptions about disk and halo contributions.
Clearly, no strong constraint can be placed on the disk luminosity function
without better statistics.
Our best estimate of $\Phi_d (M_V)$,
however, is intermediary between previous photometric and parallax surveys.
It has been shown that the former are incomplete by $\sim 20-50\%$
because of unresolved binaries (Kroupa \etal 1991, Reid 1991).
Our derived disk luminosity function based on HST data, being less affected
by this problem, consistently lies above the ground-based photometric
luminosity function shown
in the figure.
On the other hand, complete parallax surveys lead to a flatter $\Phi_d (M_V)$,
which is again consistent with unresolved binarism being an important
factor in the analysis of photometric luminosity functions (Kroupa \etal 1993).

%On the other hand, complete parallax surveys
%sample regions only \ltsima $25 pc$
%away from the solar position.
%Thus, our steeper
%(best estimate) slope of $\Phi_d (M_V)$ suggests that
%its shape may vary with scale height, in support to previous studies
%based on the dynamics of M and K dwarfs (Hawkins 1987 and Hawkins \& Bessel
%1988). The relative importance of the two effects discussed cannot
%be quatified at present, given the uncertainties in the
%normalization at $M_{br}$ and the wide range of slopes that can
%accommodate our HST counts.

\medskip\nobreak
{\bf 3.3 Comparison with models for the Galaxy}
\medskip\nobreak
The star counts for the 13 MDS fields are shown in full in figure 7.
Panel {\it a} shows the magnitude counts whereas panel {\it b} shows
the colour distribution.
Both distributions include completeness corrections and
Poissonian error bars. We also show the predicted counts from two
models for the Galaxy; the dotted lines assume only two
Galactic components (disk and halo) and correspond to a model historically
proposed by Bahcall and collaborators (Bahcall \& Soneira 1980,1984 and
references therein, hereafter BS)
whereas the solid lines include
the thick disk,
originally suggested by Gilmore \& Reid (1983), hereafter GRW.
The density laws, luminosity functions and
color-magnitude diagrams are summarized in Table~4.
They correspond to standard assumptions used by previous
star count studies (see Gilmore, Wyse \& Kuijken 1989, Gilmore, King \&
van der Kruit 1990 and Reid \& Majewski 1993 for reviews).
Notice that the density laws are the same as used in the previous section.
We should also point out that
the model predictions use $M_V(B-V)$ colour-magnitude diagrams
and V band luminosity functions. The V magnitude counts were
converted to the I band by means of several $M_V (V-I)$ relations.
The (V-I) colour distributions were obtained in the same way.
For faint and bright disk stars, fits to the $M_V(V-I)$
main sequence given
by Monet \etal (1992) and Prosser (1992) were used, respectively.
The corresponding halo
main sequence fits were taken from Bahcall \etal (1994) and
Richer \& Fahlman (1992). In the case of the GRW model,
we assumed the average between the halo and disk colour-magnitude diagrams
for the thick disk component. The contribution of red giants was obtained
by converting their (B-V) colours to (V-I), using fits to the relations
given by Prosser (1992) (for disk stars) and Brewer, Fahlman \& Richer (1993)
and Sarajedini \& Milone (1995) (for population II stars).

The two standard model predictions
for the magnitude counts are similar and
in good agreement with the observations
(figure 7a).
We derive fairly low confidence levels of inconsistency
between models and magnitude
counts, typically of $\sim 60-90\%$, depending on the magnitude range
and model considered. The BS model as shown in Table~4
fits the counts better than the GRW model
(\ltsima 80\% confidence level against \ltsima 90\%);
the latter slightly underestimates
the observed counts beyond $I \sim 22.0$.
However, the match between the GRW model and the data can be easily improved
by making only slight adjustments to the halo and thick disk parameters.
These conclusions are largely independent of
photometric calibration, since the differences between the Bahcall \etal
calibration and the standard WFPC-2 team one occur mainly
on the two extremes of the colour distribution, which do not
significantly contribute to the total magnitude counts (see figure 7b).

The model $V-I$ distributions shown
in panel {\it b} are in strong disagreement
with the data (\gtsima $99\%$ confidence level for both BS and GR).
Most of the difference
occurs in the blue region ($V-I \leq 2.0$),
where the models fail to reproduce the peak seen
at $V-I \simeq 1.5$. The discrepancies are
clearly larger than expected from the error bars and occur in a
colour range where halo stars dominate the predicted counts.
An improved match to the observed distribution can be obtained by
increasing the halo/disk local density normalization to 1/500 or making
it rounder ($c/a = 0.9$).
This later modification, however, is directly at variance with
recent measurements of the stellar halo axis ratio (Wyse \& Gilmore 1989;
Larson \& Humphries 1995), which require an axial ratio $c/a \sim 0.5$.
Should the present data require a rounder halo, then the
stellar halo must have variable flattenning, with $c/a \sim 0.5$ within
a few kpc of the plane, and a larger value beyond $\sim 10$ kpc.

A more efficient way of reproducing the observed colour distribution
is by adopting the $\omega~Cen$ $M_V(V-I)$ relation
obtained by Elson \etal (1995)
as representative of the halo stars. This later
was also derived using HST-WFPC2 and
is much steeper than the relations of
Richer \& Fahlman (1992) and Bahcall \etal (1994).
Usage of the $M_V(V-I)$ from Elson \etal leads to a much better
fit to the observed colours with only small adjustments in the
model parameters. However, given the non-uniqueness
of the CMD representative of halo stars,
plus the uncertainties in the
photometric calibration and some possible residual contamination of the
sample by non-stellar objects, any attempt to constrain the current values
of the model parameters must be considered as only tentative.

Finally, notice the reasonable agreement between the
standard BS and GRW
models over most of the colour range. The excess of
the former relatively to the latter for $V-I > 3$ is caused by the
assumption of a flat rather than declining disk luminosity function.

\bigskip
{\bf {4 CONCLUSIONS}}
\medskip\nobreak

HST field star counts were used in this paper to estimate the amount
of the missing matter in the Galaxy that could be in the form of
low-mass stars. Main sequence M dwarfs and subdwarfs do not account
for more than 25\% of the Galaxy's dark matter, unless candidates
with $M_I$ \gtsima $14.5$,
$M \sim 0.2 M_\odot$
are invoked. Such candidates would violate our current understanding
of the mass-luminosity relation for low-mass stars (Brewer \etal 1993;
Kroupa \etal 1993). Likewise, white dwarfs could be responsible for
a substantial fraction of the dark matter only if fairly large numbers of very
cool (old) such objects exist, with the additional constraint that
stars of mass lower than $M \sim 1 M_\odot$ should not
contribute to these numbers.
In a similar analysis, also using HST,
Bahcall \etal (1994) find even more stringent constraints on the amount of
dark matter in low mass stars. Those authors find no stars with
$V-I > 3.0$ and only five with $V-I > 2.0$. Their data, however, cover
only one deep WFPC-2 field, containing a total of 15 stars. Their
small dataset and preliminary photometric calibration may, at least partly,
account for the differences in the nominal upper limits to the stellar
dark matter. Notice, however, that the Bahcall \etal photometric calibration
method actually leads to an overestimate of the red star counts in our
data.

Our results are also in qualitative agreement with those
of Hu \etal (1994) and Boeshaar, Tyson \& Bernstein
(1994) although these later authors
again place stronger upper limits to the amount of stellar
dark mass than those quoted here.

The slope of the stellar luminosity function fainter than
$M_V \simeq 13$ is negative to more than
99\% confidence. Mostly because of limited statistics and
uncertainties in the conversion of
our HST V and I photometry to the Johnson-Cousins system, we are unable
to strongly constrain the slope of the disk luminosity function.
However, our best estimate for this quantity lies
between the results based on previous ground-based
photometric and volume-limited (parallax) samples.
This is consistent with unresolved stellar binarism being the cause of the
apparent inconsistency between the luminosity functions of nearby stars and of
more distant disk stars. Our inferred range of slopes disagrees with
the values derived by Richer \& Fahlman (1992) for globular clusters.
This disagreement may, at least in part, be caused by crowding of their
ground-based data. Notice that HST globular cluster star counts have
consistently also failed to reveal large numbers of low-mass stars
(Paresce \etal 1995, Elson \etal 1995).

%We tentatively suggest that the discrepancy at the faint end of
%$\Phi_d (M_V)$ found in these previous studies is
%a combination of incompleteness in the photometric
%samples (caused by unresolved binaries) and variations in the
%scale height of M dwarfs as a function of luminosity.

The current models of Galactic structure reproduce fairly well the observed
magnitude counts, with at worst slight revisions in the standard
values for their parameters being necessary.
However, there seems to be an excess of
fairly blue ($1.0 \leq V-I \leq 1.8$) stars, possibly associated
to the halo, that neither the
Bahcall \& Soneira nor the Gilmore, Reid \& Wyse models
can account for. Improved agreement can be obtained by modelling the
colour-magnitude diagram of halo stars with the
$M_V(V-I)$ relation of Elson \etal (1995), instead of using CMDs based
on ground-based data. This uncertainty in the Halo CMD, combined with
photometric calibration uncertainties and possible residual
contamination of the sample make it impossible to constrain the model
parameters with the current data. This will be considered further in
a future paper.

\bigskip
{\bf ACKNOWLEDGMENTS}
\medskip\nobreak
We thank the entire Medium Deep Survey
team, specially those at the JHU branch, for their valuable suggestions
and assistance with the MDS database. K. Glazebrook, N. Tanvir and T.
von Hippel, of IoA, Cambridge,
contributed with both software and useful suggestions.

\vfill\eject

{\bf REFERENCES}
\def\pp{\parshape 2 0truecm 13.4truecm 2.0truecm 11.4truecm}
\def\apjref #1;#2;#3;#4 {\par\pp#1, #2, #3, #4 \par}

\medskip
\hyphenation{MNRAS}
%\sevenrm

\pp Bahcall, J.N., \& Soneira, R.M., 1980, ApJS, 44, 73.

\pp Bahcall, J.N., Schmidt, M., \& Soneira, R.M., 1983, ApJ, 265, 730.

\pp Bahcall, J.N., \& Soneira, R.M., 1984, ApJS, 55, 67.

\pp Bahcall, J.N., Flynn, C., Gould, A., \& Kirhakos, S., 1994, ApJ, 435, L51.

\pp Bessel, M.S., \& Stringfellow, G.S., 1993, ARA\&A, 31, 433.

\pp Boeshaar, P.C., Tyson, J.A., \& Bernstein, G.M., 1994, in {\it Dark
Matter}, Fifth Maryland Astrophysics Conf, AIP Press.

\pp Brewer, J.P., Fahlman, G.G., \& Richer, H.B., 1993, AJ, 105, 2158.

\pp Carr, B., 1994, ARA\&A, 32, 531.

\pp Dahn, C., Wielen, R., Liebert, J., \& Harrington, R.S., 1986, AJ, 91, 621.

\pp Elson, R.A.W., Gilmore, G., Santiago, B.X., \& Casertano, S., 1995, AJ,
in press.

\pp Faber, S.M., \& Gallagher, J.S., 1979, ARA\&A, 17, 135.

\pp Fich, M., \& Tremaine, S., 1991, ARA\&A, 29, 409.

\pp Gilmore, G., \& Reid, N., 1983, MNRAS, 202, 1025.

\pp Gilmore, G., 1984, MNRAS, 207, 223.

\pp Gilmore, G., Wyse, R., \& Kuijken, K., 1989, ARA\&A, 27, 555.

\pp Gilmore, G., King, I.R., \& van der Kruit, P.C., 1990, {\it The Milky
Way as a Galaxy}, University Science Books, CA.

\pp Griffiths, R., \etal, 1994, ApJ, 435, L19..

\pp Jahreiss, H., 1987, in {\it Impacts des Surveys du visible sur notre
Connaissance de La Galaxie}, ed: A. Fresneau, M. Hamon, p73.
Strasbourg: Comptes Rendus Journ. Strasbourg 9\`eme Reunion.

\pp Harris, H.C., Baum, W.A., Hunter, D.A., \& Kreidl, T.J., 1991, AJ, 101,
677.

\pp Hawkins, M.R.S., 1987, MNRAS, 223, 845.

\pp Hawkins, M.R.S., \& Bessel, M.S., 1988, MNRAS, 234, 177.

\pp Holtzmann, J.A., \etal, 1995, PASP, 107, 156.

\pp Hu, E.M., Huang, J.S., Gilmore, G., \& Cowie, L.L., 1994, Nat, 371, 493.

\pp Kron, R.G., 1980, ApJS, 43, 305.

\pp Kroupa, P., Tout, C.A., Gilmore, G., 1991, MNRAS, 251, 293.

\pp Kroupa, P., Tout, C.A., Gilmore, G., 1993, MNRAS, 262, 545.

\pp Larsen, J.A., \& Humphries, R.M., 1994, ApJ, 436, L139.

\pp Larson, R., 1986, MNRAS, 218, 409.

\pp Liebert, J., 1980, ARA\&A, 18, 63.

\pp Liebert, J., Dahn, C.C., \& Monet, D.G., 1988, ApJ, 332, 891.

% \pp Majewski, S.R., 1993, ARA\&A, 31, 575.

\pp Mathews, G.J., Schramm, D.N., \& Meyer, B.S., 1993, ApJ, 404, 476.

\pp Monet, D.G., Dahn, C.C., Vrba, F.J., Harris, H.C., Pier, J.R., \etal
1992, AJ, 103, 638.

\pp Ojha, D.K., Bienaym\'e, O., Robin, A.C., Mohan, U., 1994, A\&A, 290, 771.

\pp Paresce, F., De Marchi, G., \& Romaniello, M., 1995, ApJ, 440, 216.

\pp Perrin, M.N., Friel, E.D., Bienaym\'e, O., Cayrel, R., Barbuy, B.,
\& Baulon, J., 1995, A\&A, in press.

\pp Prosser, C., 1992, AJ, 103, 488.

\pp Reid, I.N., \& Gilmore, G.F., 1982, MNRAS, 201, 73.

\pp Reid, I.N. 1990, MNRAS, 247, 70.

\pp Reid, I.N., 1991, AJ, 102, 1428.

\pp Reid, I.N., 1992, in {\it Galaxy Evolution: The Milky Way Perspective},
ed: S. Majewski, ASP Conf. Ser., vol 49, p37.

\pp Reid, I.N., \& Majewski, S.R., 1993, ApJ, 409, 635.

\pp Richer, H.R., \& Fahlman, G.G., Buonanno, R., Fusi Pecci, F., Searle, L.,
\& Thompson, I., 1991, ApJ, 381, 147.

\pp Richer, H.R., \& Fahlman, G.G., 1992, Nat, 358, 383.

\pp Rubin, V.C., Ford, W.K., \& Thonnard, N., 1982, ApJ, 261, 439.

\pp Sarajedini, A., \& Milone, A.A., 1995, AJ, 109, 269.

\pp Silk, J., 1990, in {\it Baryonic Dark Matter}, eds: Lynden-Bell, D.,
\& Gilmore, G. Dordrecht: Kluwer.

\pp Soubiran, C., in {\it Astronomy from Wide-Field Imaging}, IAU Symp. 161,
ed H. McGillivray, 1993.

\pp Stobie, R.S., \& Ishida, K., 1987, AJ, 93, 624.

\pp Stobie, R.S., Ishida, K., Peacock, J.A.,, 1989, MNRAS, 238, 709.

\pp Tamanaha, C.M., Silk, J., Wood, M.A., \& Winget, D.E., 1990, ApJ, 358,
164.

\pp von Hippel, T., Gilmore, G., \& Jones, D.H.P., 1995, MNRAS, 273, L39.

\pp Walker, T., Steigman, G., Schrammn, D.N., Olive, K.A., Kang., H.S.,
1991, ApJ, 376, 51.

\pp Wielen, R., Jahreiss, H., \&, Kr\"oger, R., 1983, in {\it Nearby Stars
and the Stellar Luminosity Function, IAU Colloq 76}, eds: A.G. Davis
Philip, A.R., Upgren, p163. Schenectady, NY: L. Davis.

%\pp Wielen, R., 1974, in {\it Highlights of Astronomy 3}, ed: G. Contopoulos,
%p489. Dordrecht: Reidel.

\pp Wood, M.A., 1992, ApJ, 386, 539.

\pp Wyse, R.F.G., \& Gilmore, G., 1989, {\it Comments on Astrophysics},
13, 135.

\vfill\eject

\centerline{{\bf Table~1}.
MDS fields}
\vskip 0.5 true cm
\hrule
\vskip 0.2 true cm
\tabskip=1em plus2em minus.5em
\halign to\hsize
{#\hfil&&\hfil#&\hfil#&\hfil#&\hfil#&\hfil#&\hfil#\cr
Field & l & b & $t_I$ & $t_V$ & \#I & \#V \cr
\noalign{\smallskip\hrule\smallskip}
ucs0 & 233.664 & -62.412  & 6600 &  1500 &  3 & 2 \cr
uy40 &  33.912 &  66.759  & 6000 &  5200 &  6 & 6 \cr
ubi1 & 133.943 & -64.934  & 6300 &  3300 &  3 & 2 \cr
ut20 &  15.986 &  39.950  & 6000 &  5200 &  6 & 6 \cr
ux40 &  35.789 &  56.483  & 7500 &  3300 &  4 & 2 \cr
uad0 &  83.864 & -76.396  & 2000 &  1200 &  2 & 2 \cr
usa0 &  56.733 &  34.209  & 6300 &  5400 &  3 & 3 \cr
ua-0 & 215.003 & -87.548  & 4200 &  2100 &  2 & 2 \cr
uy41 &  34.347 &  66.652  & 8000 &  2400 &  8 & 4 \cr
uj70 & 326.375 & -29.580  & 6300 &  5400 &  3 & 3 \cr
ut21 &  16.221 &  40.059  & 12000 &  3600 & 12 & 6 \cr
uqa0 &  52.025 &  27.812  & 2280 &  1200 &  4 & 2 \cr
ux41 &  35.583 &  56.427  & 6000 &  3300 &  3 & 2 \cr
}
\vskip 0.2 true cm
\hrule

\vfill\eject

 \centerline{{\bf Table~2}.
Constraints on the slope of the halo V band luminosity function.}
\smallskip
\vskip 0.2 true cm
\hrule
\vskip 0.2 true cm
\tabskip=1em plus2em minus.5em
\halign to\hsize
{#\hfil&\hfil#&\hfil#&\hfil#\cr
Nature of constraint & Domain in $M_V$ & Mass & $\gamma_h$ \cr
\noalign{\smallskip\hrule\smallskip}
Stellar DM & $13.0 - 19.0$ & Kroupa etal & 1.56 \cr
Stellar DM & $13.0 - 19.0$ & 0.2 $M_\odot$ & 1.35 \cr}
{\bigskip\hrule\bigskip}
\halign to\hsize
{#\hfil&\hfil#&\hfil#&\hfil#\cr
Nature of constraint & Domain in $M_V$ & $\gamma_h$ & DM Fractions \cr
\noalign{\smallskip\hrule\smallskip}
MDS counts & $13.0 - 19.0$ & 1.19 & 0.23/0.55 \cr}
\vskip 0.2 true cm
\hrule

\vfill\eject

\centerline{{\bf Table~3}.
The faint end of the stellar luminosity function from star counts}
\vskip 0.5 true cm
\hrule
\vskip 0.2 true cm
\tabskip=1em plus2em minus.5em
\halign to\hsize
{#\hfil&&\hfil#&\hfil#\cr
Components & \# of stars & $\gamma$ \cr
\noalign{\smallskip\hrule\smallskip}
Disk+Halo & 39 & $-0.29$ \cr
Disk & 39 & $-0.26$ \cr
Disk+Halo & 54 (99\%) & $-0.12$ \cr
Disk & 54 (99\%) & $-0.09$ \cr}
\vskip 0.2 true cm
\hrule

\vfill\eject

\centerline{{\bf Table~4}.
Models for the Galaxy.}
\vskip 0.2 true cm
\hrule
\vskip 0.2 true cm
\tabskip=1em plus2em minus.5em
\halign to\hsize
{#\hfil&&\hfil#&\hfil#\cr
{}~~~~~~~~ & Bahcall \& Soneira & Gilmore, Reid \& Wyse \cr
\noalign{\smallskip\hrule\smallskip}
\noalign{\centerline {Thin Disk}}
Scale length (pc) & 3500 & 3500 \cr
Scale height (pc) & 90-325 & 90-325 \cr
C-M diagram & Johnson 1965 & M67 \cr
Lum. Function & Wielen & Wielen, Gilmore \& Reid \cr
\noalign{\smallskip\hrule\smallskip}
\noalign{\centerline {Thick Disk}}
Scale length (pc) & ~~~~~~ & 3500 \cr
Scale height (pc) & ~~~~~~ & 1300 \cr
C-M diagram & ~~~~~~~ & 47 Tuc \cr
Lum. Function & ~~~~~~ & 47 Tuc \cr
Local normalization & ~~~~~~ & 1/50 \cr
\noalign{\smallskip\hrule\smallskip}
\noalign{\centerline {Halo}}
de Vaucouleurs radius (pc) & 2670 & 2700 \cr
C-M diagram & M13 & M 5 \cr
Lum. Function & Wielen & 47 Tuc \cr
Local normalization & 1/500 & 1/800 \cr
Axis ratio & 0.80 & 0.80 \cr}

\vfill\eject

\def\fig #1, #2, #3, #4 {\smallskip\leftskip2.5em \parindent=0pt
#4}
%The following macro will give inserted figures. Comment it out if you
%don't have psfig.
\input psfig
\def\fig #1, #2, #3, #4 {
\topinsert
\smallskip
\centerline{\psfig{figure=#1,height=#2 in,width=#3 in}}
\medskip
{\smallskip\leftskip2.5em \parindent=0pt
#4}
\endinsert}

\centerline {\bf FIGURE CAPTIONS}

\fig 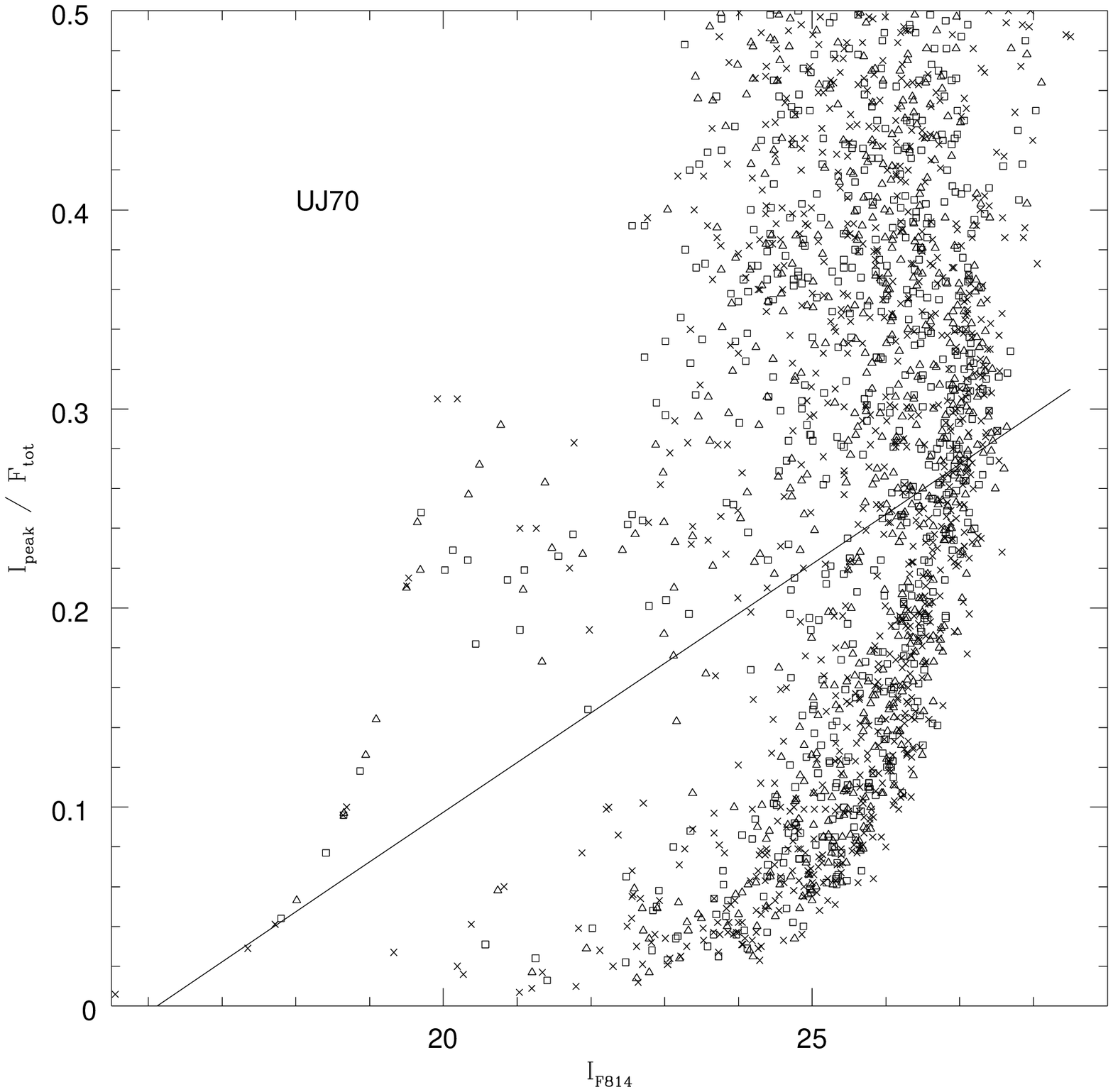, 5, 5, {1- The $I_{peak} / F_{tot} \vs I_{F814}$ relation
for one of the
MDS fields (UJ70). Objects detected in each WFC chip are shown with
different symbols: open triangles (chip 2), open squares (chip 3) and
crosses (chip 4). The solid line represents the star-galaxy separator used
in this work. }

\fig 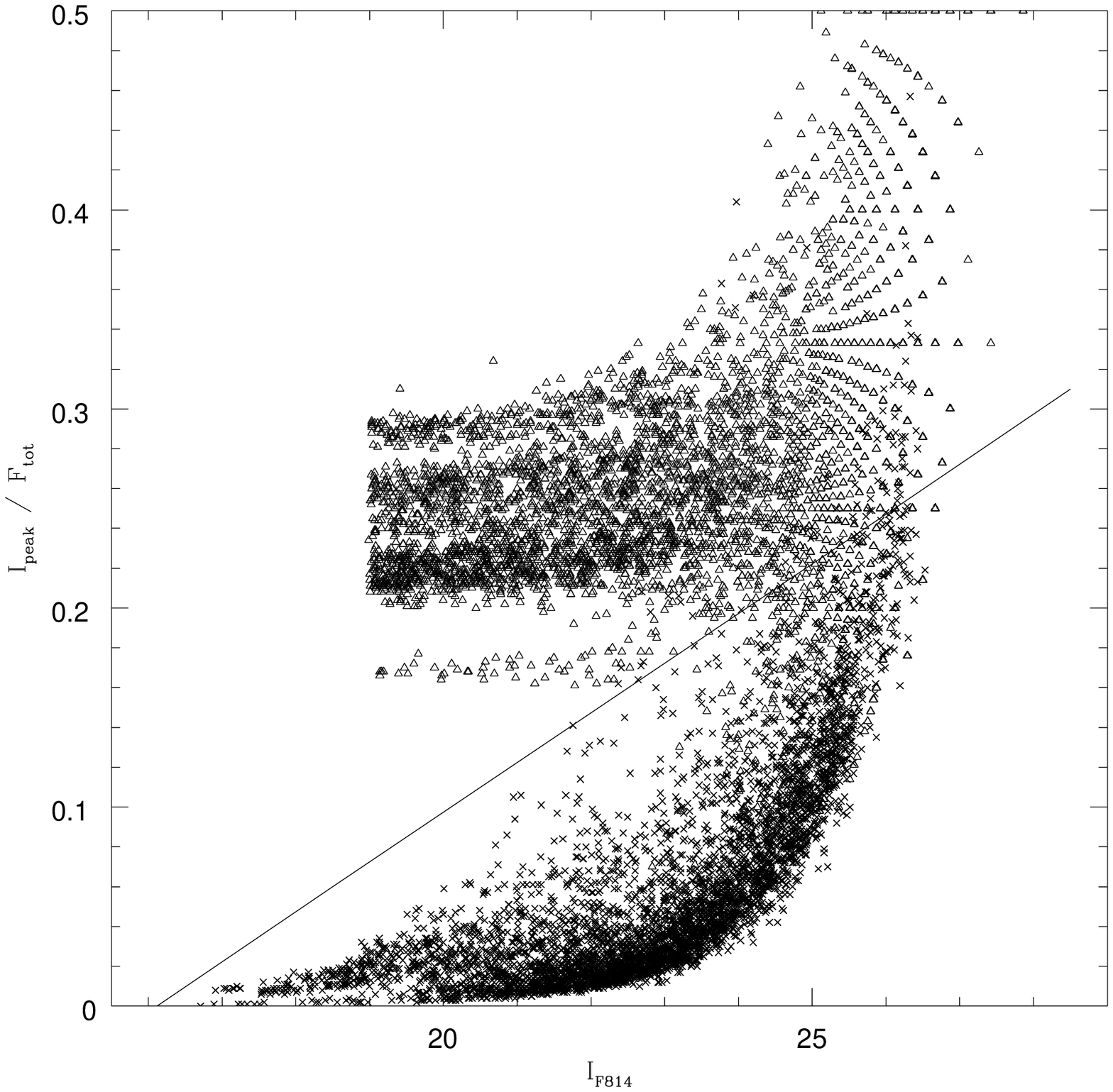, 5, 5, {2- The same plot as in figure 1 but
now for a simulation of
several MDS images containing thousands of stars and galaxies. }

\fig 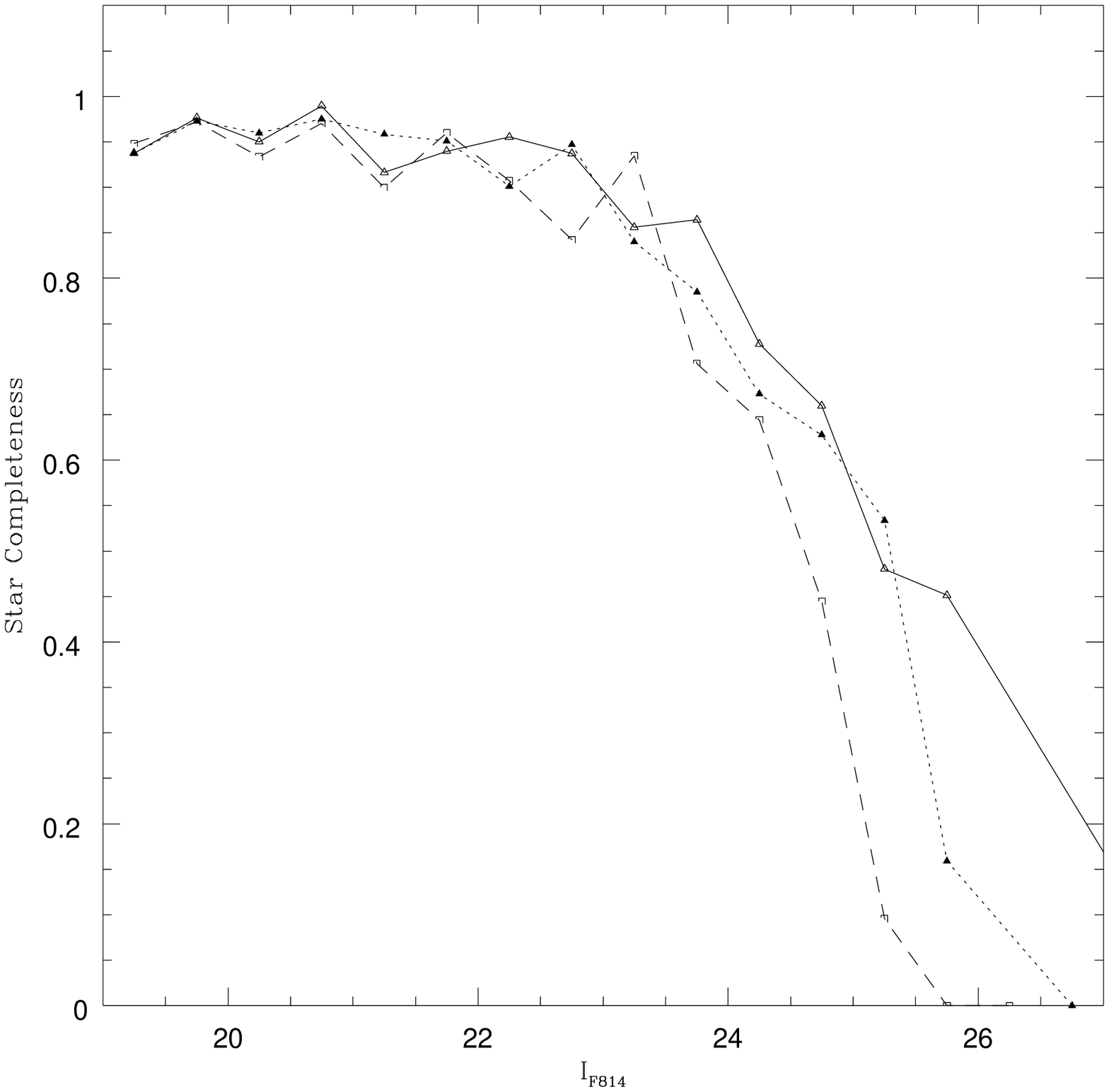, 5, 5, {3- Star completeness function for
three different simulations,
with decreasing S/N. The S/N for a $I = 24 mag$ point source are
as follows: solid line and open triangles, 30; dotted line and solid
triangle, 25; dashed line and open square, 15. }

\fig 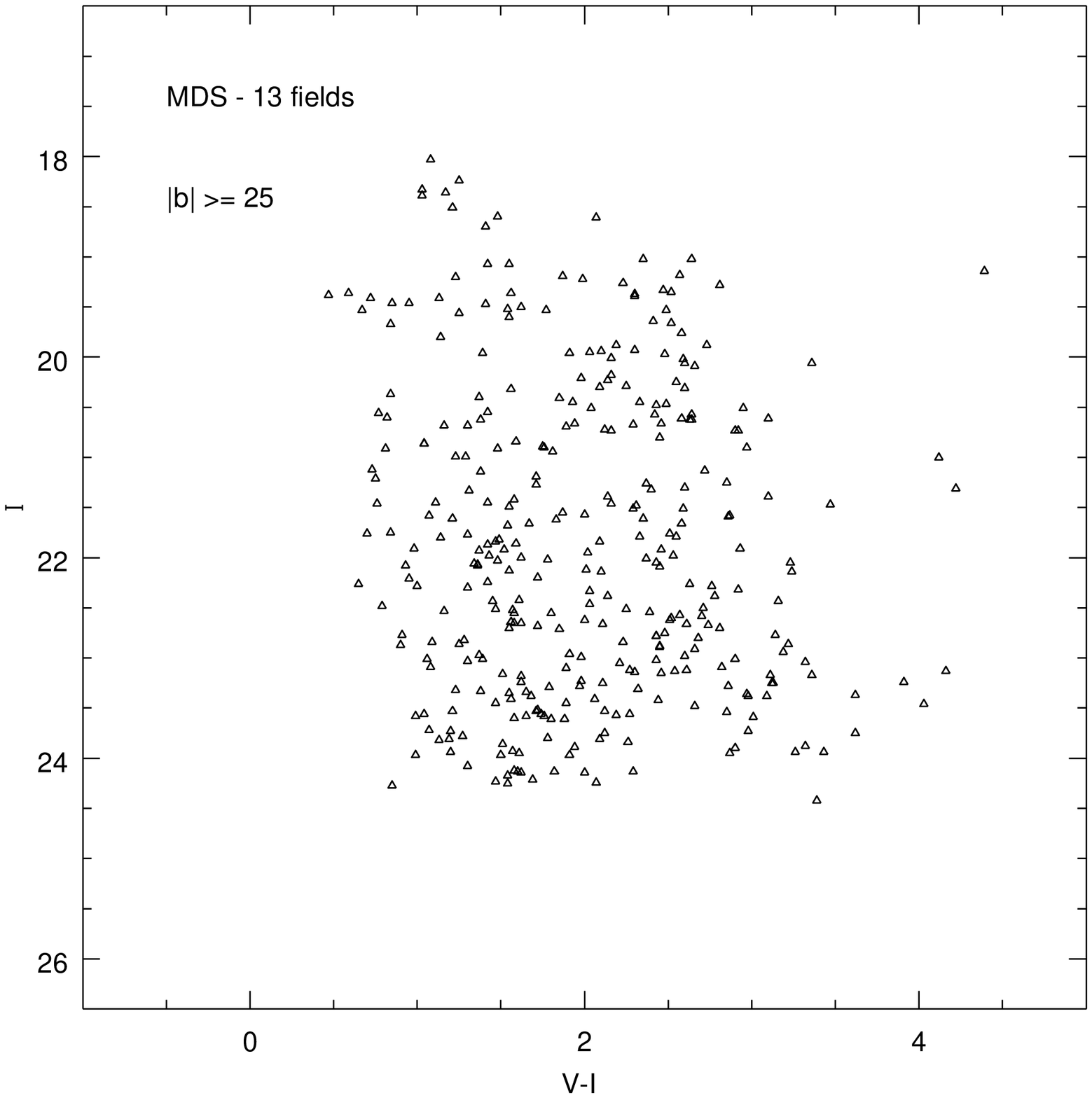, 5, 5, {4- Colour-magnitude diagram
for the stellar sample selected
{}from the 13 MDS fields. Both magnitudes and colours are in the
Johnson-Cousins system. Conversion from the HST passbands was done using
the transformations quoted in Harris \etal (1991) and Holtzmann \etal
(1995). }

\fig 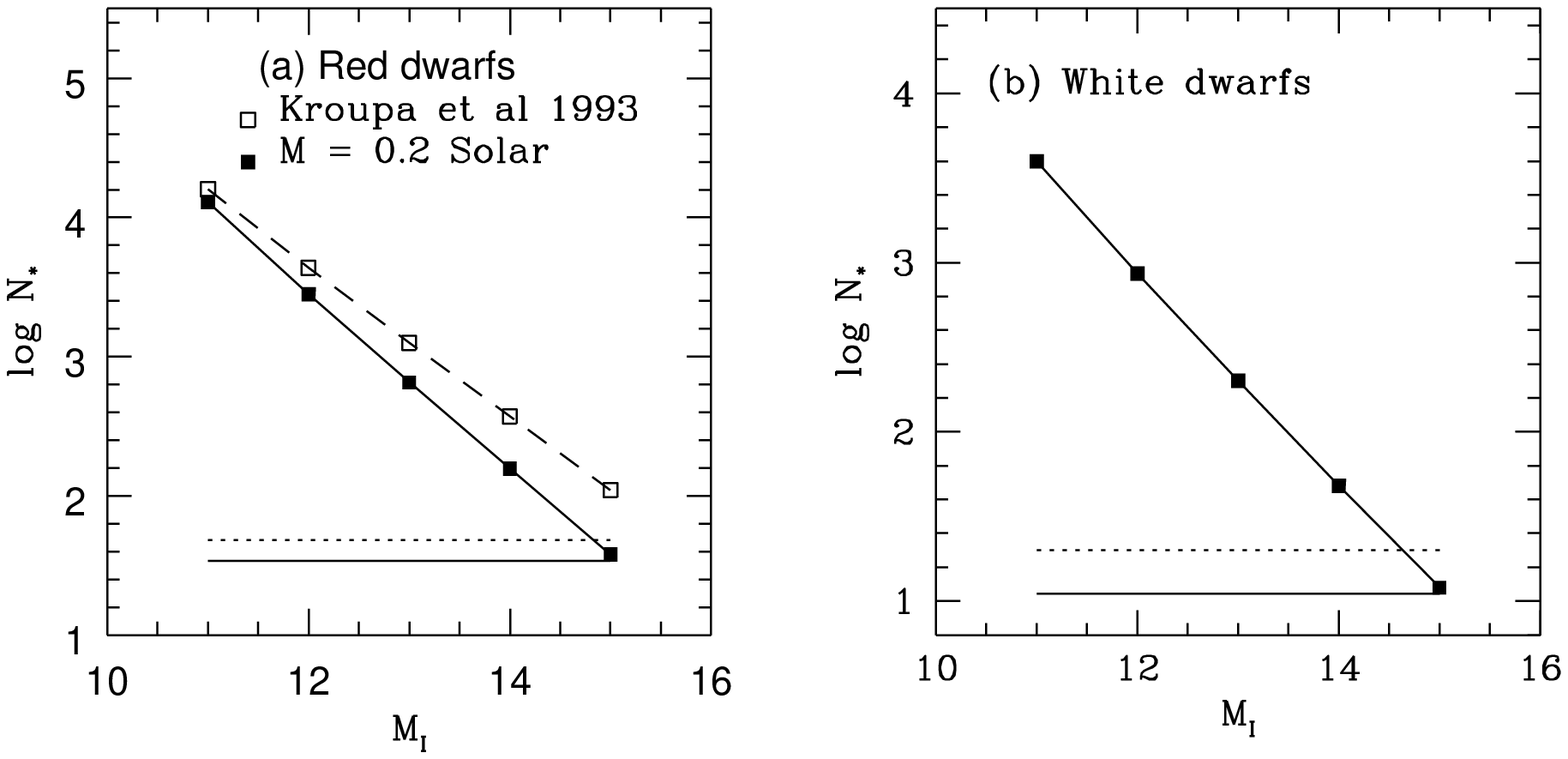, 5, 5, {5- {\it a)} Predicted number of M dwarfs and subdwarfs
with different absolute magnitudes $M_I$ and Masses M. Masses
for the open
squares and dashed line were taken from Kroupa \etal (1993) $M_V - M$
relation, assuming a $V-I = 3.0$ for all candidates. The solid symbols
and line represent predictions for a fixed mass of 0.2 $M_\odot$.
The horizontal solid lines correspond to the observed number of MDS stars
with $V-I \geq 3.0$ (solid line) and its 99\% Poisson deviate (dotted).
{\it b)} Predicted number of 0.65 $M_\odot$ white dwarfs of different
$M_I$ magnitudes. The horizontal lines are the observed
number of $V-I \leq 0.8$ stars (solid) and its corresponding
99\% Poisson deviate (dotted). }

\fig 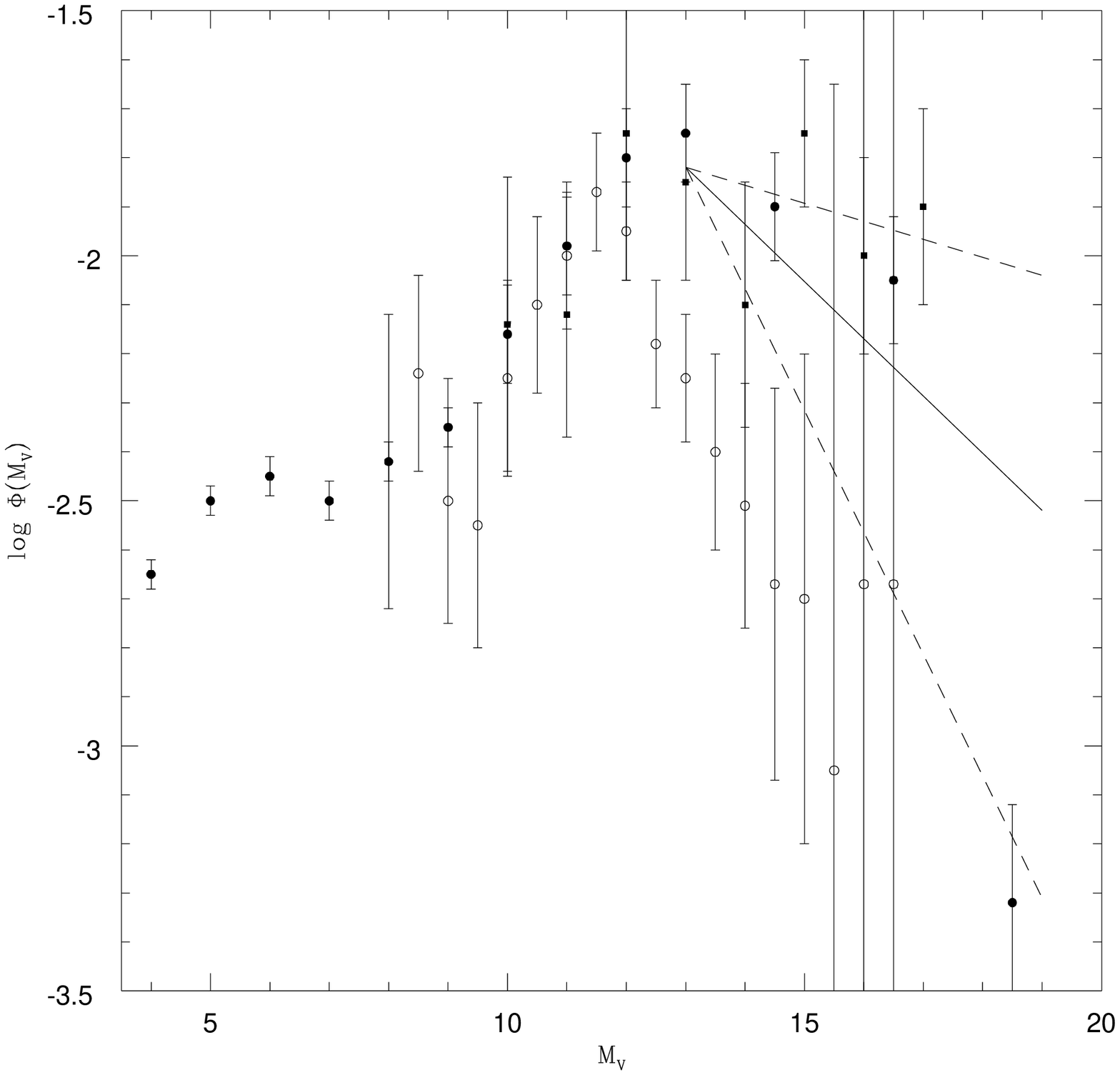, 5, 5, {6- Solid line: the best estimate of the disk luminosity
function for $M_V \geq 13$, obtained by matching
standard model predictions of disk and halo to the observed
number counts of red stars. Upper dashed line: an upper limit to the
disk luminosity function, utilizing the 99\% Poisson deviate
of the observed number and assuming that only disk stars contribute
to this number. Lower dashed line: a lower limit
to $\Phi_d (M_V)$, utilizing the 1\% Poisson deviate of the
MDS counts and considering both disk and halo contributions.
The points represent results from previous works, taken from
figure 2 of Bessel \& Stringfellow (1993): filled circles,
Wielen \etal (1983) and Jahreiss (1987); filled squares, Dahn \etal
(1986); open circles, Stobie \etal (1989). }

\fig 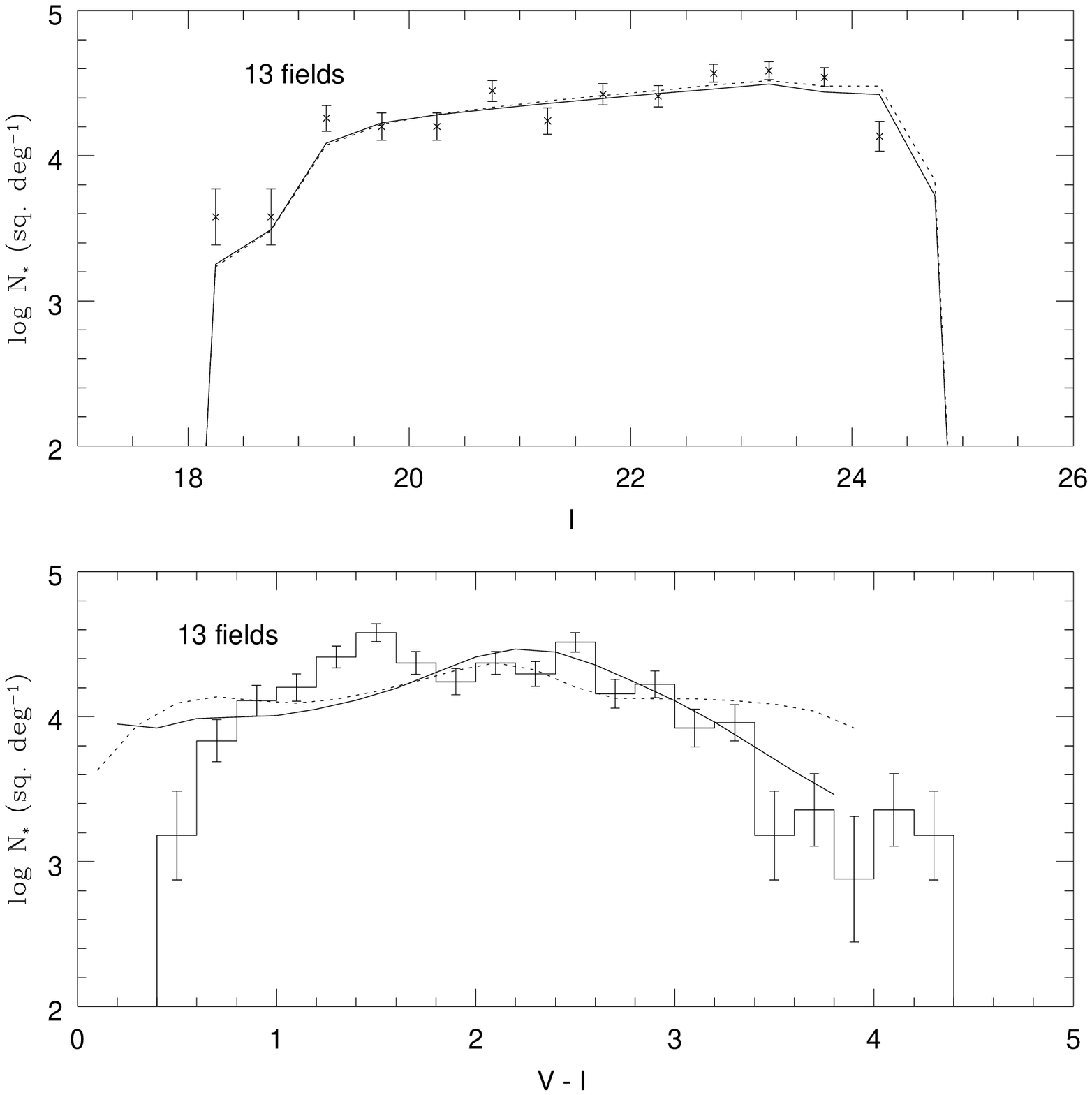, 5, 5, {7-  {\it a)} The points represent I band
magnitude counts
obtained from the MDS data. Error bars are Poissonian.
The solid (dotted) line corresponds to the Gilmore, Reid \& Wyse
(Bahcall \& Soneira) standard model prediction.
{\it b)} Observed and predicted colour distributions for the
13 MDS fields. The histograms correspond to the
data with associated Poisson error bars.
The solid and dotted lines correspond to
model predictions using the same convention as in panel {\it a}. }

\vfill\eject
\end